\documentclass[preprint,12pt]{elsarticle}

\usepackage{caption}
\usepackage{subcaption}
\usepackage{amssymb}
\usepackage{amsmath}
\usepackage{amsfonts}
\usepackage{makecell}
\usepackage{textcomp}
\usepackage{subcaption}
\usepackage{graphicx}
\usepackage{rotating}
\usepackage{makecell}

\journal{Economic analysis and policy}

\begin{document}

\begin{frontmatter}








\author[1]{Kunpeng Wang\corref{cor1}}
\ead{kunpeng.wang@scu.edu.cn}
\author[2]{Jiahui Hu}
\ead{Jiahui.Hu22@student.xjtlu.edu.cn}

\address[1]{Sichuan University-Pittsburgh Institute, Sichuan University, Chengdu, Sichuan Province, 610207, China}
\address[2]{International Business School Suzhou, Xi’an Jiaotong-Liverpool University, Suzhou, Jiangsu Province, 215123, China}



\title{Governance of Technological Transition: A Predator-Prey Analysis of AI Capital in China's Economy and Its Policy Implications}


\begin{abstract}
The rapid integration of Artificial Intelligence (AI) into China's economy presents a classic governance challenge: how to harness its growth potential while managing its disruptive effects on traditional capital and labor markets. This study addresses this policy dilemma by modeling the dynamic interactions between AI capital, physical capital, and labor within a Lotka–Volterra predator-prey framework. Using annual Chinese data (2016–2023), we quantify the interaction strengths, identify stable equilibria, and perform a global sensitivity analysis. Our results reveal a consistent pattern where AI capital acts as the ‘prey’, stimulating both physical capital accumulation and labor compensation (wage bill), while facing only weak constraining feedback. The equilibrium points are stable nodes, indicating a policy-mediated convergence path rather than volatile cycles. Critically, the sensitivity analysis shows that the labor market equilibrium is overwhelmingly driven by AI-related parameters, whereas the physical capital equilibrium is also influenced by its own saturation dynamics. These findings provide a systemic, quantitative basis for policymakers: (1) to calibrate AI promotion policies by recognizing the asymmetric leverage points in capital vs. labor markets; (2) to anticipate and mitigate structural rigidities that may arise from current regulatory settings; and (3) to prioritize interventions that foster complementary growth between AI and traditional economic structures while ensuring broad-based distribution of technological gains.
\end{abstract}


\end{frontmatter}


\section{Introduction}
\label{s1}

The rapid integration of artificial intelligence (AI) into national economies presents a critical challenge for policymakers worldwide: how to harness AI's growth potential while mitigating its disruptive effects on established capital and labor markets. China, with its state-led development model and ambitious AI strategies, offers a compelling case study for examining this policy dilemma \cite{zhu2018}. While AI is widely recognized as a driver of productivity \cite{ai2}, its long-term interactions with traditional factors of production remain empirically contested and theoretically ambiguous. Some studies suggest AI complements physical capital and augments labor \cite{he2019, ai1}, whereas others argue it may substitute for both, leading to capital shallowing and labor displacement \cite{jones}. These conflicting predictions create significant uncertainty for industrial policy, investment guidance, and labor-market regulation.

A key limitation of conventional economic analysis lies in its treatment of AI primarily as a component of exogenous or endogenous technical change within neoclassical frameworks \cite{ai2}. This approach often fails to capture the dynamic, nonlinear feedback that characterizes the real-world co-evolution of AI with physical capital and labor. If AI operates as a distinct factor of production with self-reinforcing capabilities \cite{lu2021}, its expansion may trigger complex resource reallocations across sectors, altering not just the levels but the functional relationships between economic inputs over time. Understanding these dynamic interdependencies is essential for designing policies that promote stable, inclusive growth during technological transition.

Recent literature highlights this complexity. Aghion et al. \cite{jones}, modeling AI as automation within a Zeira-type framework \cite{zeira1998}, find it to be labor-augmenting yet capital-depleting—a counterintuitive result that underscores non-trivial rebalancing. Huang \cite{huang2024}, using a task-based model \cite{ace2018}, shows that AI adoption triggers cross-industry capital and labor flows mediated by price effects. These findings collectively suggest that treating AI as an independent factor and modeling its interactions dynamically is crucial for accurate policy assessment \cite{lu2021}.

To address this gap, we introduce a coupled-system perspective from ecological economics. We model the pairwise relationships between AI capital, physical capital, and labor in China using the generalized Lotka–Volterra (predator–prey) framework. This method allows us to quantify not only the direction (positive/negative) but also the strength and asymmetry of the interactions, conceptualizing the economy as an ecosystem where each factor's growth depends on the state of the others. The Lotka–Volterra model has proven effective in analyzing competitive dynamics in technological and industrial contexts \cite{Lee2005, wu2012, chiang2012, xia2022}, making it suitable for this investigation.

Using annual Chinese data (2016–2023), we estimate the interaction parameters for both the AI–physical capital and AI–labor subsystems. We then conduct equilibrium and stability analyses to identify the long-run steady states implied by the estimated dynamics. Finally, we perform a global sensitivity analysis (Sobol' indices) to determine which parameters most critically influence these equilibrium outcomes—information vital for assessing policy leverage and system robustness.

Our analysis yields three policy-relevant findings. First, we identify a stable predator–prey dynamic in both subsystems, with AI capital acting as the "prey" that stimulates physical capital accumulation and labor compensation, while facing only weak constraining feedback. Second, the equilibrium points are stable nodes, indicating monotonic convergence shaped by current policy settings rather than endogenous cycles. Third, sensitivity analysis reveals an asymmetry: labor market outcomes are overwhelmingly driven by AI-side parameters, whereas physical capital outcomes are also influenced by its own saturation dynamics. These results provide a quantitative, systemic foundation for designing AI integration policies that balance innovation with stability.

The remainder of the paper is organized as follows. Section \ref{s2} presents the Lotka–Volterra modeling framework. Section \ref{s3} describes the data and empirical estimation. Section \ref{s4} analyzes the equilibrium and stability properties. Section \ref{s5} reports the global sensitivity analysis. Section \ref{s6} concludes with targeted policy implications and avenues for future research.

\section{Model formulation}
\label{s2}

To analyze the policy-relevant dynamics of co-evolving economic factors, we employ the Lotka-Volterra model, a framework originally developed for competing biological species but increasingly applied to economic systems \emph{etc} \cite{Lee2005,wu2012,chiang2012,xia2022} to capture nonlinear feedback and interdependence. The general form of the Lotka-Volterra model with two species is as follows 
\begin{align}
& \frac{d x}{d t}=a_1 x(t)+b_{11} x(t)^2+b_{12} x(t) y(t),  \label{lveqn1} \\
& \frac{d y}{d t}=a_2 y(t)+b_{21} y(t)x(t)+b_{22} y(t)^2,
\label{lveqn2}
\end{align}
where $a_i$ is the growth parameter for the species $i$, $b_{ii}$ represents the carrying capacity parameter for the species $i$ and $b_{ij}$, $i \neq j$ stands for the interaction parameter for the species $i$. In particular, we can determine the type of interaction as shown in Table \ref{table1}.


\begin{table}[h!]
\centering
\begin{tabular}{|l|l|l|l|}
\hline
\( b_{12} \) & \( b_{21} \) & Type & Explanation \\
\hline
\(+\) & \(+\) & Pure competition & \makecell[l]{Both species negatively affect \\ each other's growth or survival.} \\
\hline
\(-\) & \(-\) & Mutualism & \makecell[l]{Both species benefit from \\ each other's existence \\ (symbiosis or win-win).} \\
\hline
\(+/-\) & \(-/+\) & Predator-Prey & \makecell[l]{One species (predator) benefits \\ at the expense of the other (prey).} \\
\hline
\(+/0\) & \(0/+\) & Amensalism & \makecell[l]{One species is negatively affected, \\ while the other is unaffected.} \\
\hline
\(-/0\) & \(0/-\) & Commensalism & \makecell[l]{One species benefits, \\ while the other is unaffected.} \\
\hline
\(0\) & \(0\) & Neutralism & \makecell[l]{No interaction \\ between the species.} \\
\hline
\end{tabular}
\caption{The signs of \( b_{12} \) and \( b_{21} \) determine the type of species interaction.}
\label{table1}
\end{table}

Leslie \cite{leslie1958} showed that  the equations \eqref{lveqn1} and \eqref{lveqn2} can be presented in the following discrete form to be compared with the data in discrete time  (see more applications in \cite{Lee2005,wu2012,chiang2012,xia2022})
\begin{align}
x(k+1) & =\frac{\alpha_1 x(k)}{1-\beta_1 x(k)-\gamma_1 y(k)}, \label{discrete1} \\
y(k+1) & =\frac{\alpha_2 y(k)}{1-\beta_2 y(k)-\gamma_2 x(k)},  \label{discrete2}
\end{align}
where $a_i= \ln \alpha_i$, $b_{ii}=\dfrac{\beta_i \ln \alpha_i}{\alpha_i-1}$ and $b_{ij}=\dfrac{\gamma_i \ln \alpha_i}{\alpha_i-1}$, $i,j=1,2$, $i \neq j$ and  $k=1,2, \ldots, n-1$, the index $n$ denotes the size of the data. Note that $\alpha_i$ and $\beta_i$ are logistic parameters for $x$ and $y$ when there is no other species influencing the living of species $i$ while $\gamma_i$ represents the interaction effect that each species exerts on the other. Also, the parameters in the transformation form are well-defined if  $\alpha_i>0$ and $\alpha_i \neq 1$. 

The above equations \eqref{discrete1} and \eqref{discrete2} can be further rewritten in the following form for regression analysis 
\begin{align}
\frac{x(k)}{x(k+1)} = \alpha'_1+\beta_1'x(k)+\gamma_1' y(k), \label{fit1} \\ 
\frac{y(k)}{y(k+1)} = \alpha'_2+\beta_2'x(k)+\gamma_2' y(k), \label{fit2}
\end{align}
where $\alpha'_i =\dfrac{1}{\alpha_i}$, $\beta_i'=-\dfrac{\beta_i}{\alpha_i}$ and $\gamma_i'=-\dfrac{\gamma_i}{\alpha_i}$, $i=1,2$.


\section{The empirical simulation of the model}
\label{s3}
The Chinese AI market is one of the largest and fastest-growing in the world, driven by strong government support, corporate investment, and widespread adoption across industries and integration of industry and academia \cite{zhu2018}. China provides a critical case study due to its active, state-led approach to AI governance, making it an ideal setting to examine how policy shapes the dynamic integration of a new technological factor with traditional economic structures. AI market size is proxied by funding raised by AI companies.  The AI market size used in this paper considers 6 different sectors in AI including AI Robotics, Autonomous \& Sensor Technology, Computer Vision, Machine Learning, Natural Language Processing, and Generative AI. We assume that the interactions among AI capital, physical capital and labor can be represented by a Lotka-Volterra model. 

AI capital is measured by the AI market size \cite{askci2024,ai2025} while physical capital is gauged by the total investment in fixed assets \cite{nbsc2023}. References \cite{nbsc2023, nbsc2024, wage2024} provide data on the average wages of urban employees and the urban employed population. The total labor is calculated by multiplying the average wage by the employed population.  All data sets are the most up-to-date from available years 2016 to 2023  as shown in Table \ref{table2}.  

\begin{table}[htbp]
\centering
\begin{tabular}{|l|r|r|r|}
\hline 
Year & \begin{tabular}[c]{@{}c@{}}AI Capital\\ (Unit: billion yuan)\end{tabular} & \begin{tabular}[c]{@{}c@{}}Physical Capital\\ (Unit: billion yuan)\end{tabular} & \begin{tabular}[c]{@{}c@{}}Labor\\ (Unit: billion yuan)\end{tabular} \\
\hline 
2016 & 15.40 & 37202.10 & 22770
 \\
\hline 
2017 & 31.80 & 39492.60 & 25500
\\
\hline 
2018 & 59.30 & 41821.50 & 28210
 \\
\hline 
2019 & 93.60 & 43954.10 & 31820
 \\
\hline 
2020 & 138.90 & 45115.50 & 34880
 \\
\hline 
2021 & 162.10 & 47300.30 & 39700
\\
\hline 
2022 & 170.60 & 49596.60 & 42390
 \\
\hline 
2023 & 213.70 & 50970.80 & 44650
 \\
\hline 
\end{tabular}
\caption{AI and physical capital data in China from 2016 to 2023.}
\label{table2}
\end{table}

\begin{table}[htbp]
\centering
\begin{tabular}{|c|c|c|}
\hline
\textbf{Parameter} & \textbf{AI Capital vs Physical Capital} & \textbf{AI Capital vs Labor} \\
\hline
\multicolumn{3}{|c|}{\textbf{Equations \eqref{fit1} and \eqref{fit2}}}  \\ 
\hline
$\alpha_1'$       & 0.021224       & 0.023710 \\
$\beta_1'$        & 0.001769       & 0.000246 \\
$\gamma_1'$       & 0.000012       & 0.000021 \\
Adjusted $R^2$    & 0.9908         &  0.9909 \\
\hline
$\alpha_2'$       & 0.007191       & 0.011324 \\
$\beta_2'$        & -0.001578      & -0.004431 \\
$\gamma_2'$       & 0.000025       & 0.000041 \\
Adjusted $R^2$    & 0.9995         &  0.9989 \\
\hline
\multicolumn{3}{|c|}{\textbf{Equations \eqref{discrete1} and \eqref{discrete2}}} \\ 
\hline
$\alpha_1$        & 47.1160        & 42.1757 \\
$\beta_1$         & -0.08337       & -0.010375 \\
$\gamma_1$        & -0.000578      & -0.000888 \\
$\alpha_2$        & 139.0605       & 88.3049 \\
$\beta_2$         & 0.219529       & 0.391303 \\
$\gamma_2$        & -0.003539      & -0.003656 \\
\hline
\multicolumn{3}{|c|}{\textbf{Equations \eqref{lveqn1} and \eqref{lveqn2}}} \\ 
\hline
$a_1$             & 3.852613       & 3.741844\\
$b_{11}$          & -0.006965      & -0.000943 \\
$b_{12}$          & -0.000048      & -0.000081 \\
$a_2$             & 4.934909       & 4.480796 \\
$b_{21}$          & 0.007846       & 0.020083 \\
$b_{22}$          & -0.000126      & -0.000187 \\
\hline
\end{tabular}
\caption{Estimated values of parameters and goodness of fit.}
\label{table3}
\end{table}

%
Notice that $\alpha'_i = \frac{1}{\alpha_i}$ is assumed to be positive and not equal to 1 in equations \eqref{fit1} and \eqref{fit2}. Therefore, we use the zero-intercept regression method \cite{eisen2003} for model fitting. The intercept $\alpha_i'$ is treated as a post-hoc parameter and evaluated based on the difference between the fitted and empirical values of the ratios $\frac{x(k)}{x(k+1)}$ and $\frac{y(k)}{y(k+1)}$, respectively. The models \eqref{fit1} and \eqref{fit2} compare well with the data presented in Table \ref{table2}, with the detailed regression results shown in Table \ref{table3}. Additionally, the original coefficient values are derived and presented in Table \ref{table3}.

\begin{figure}[htbp]
    \centering 
    \begin{subfigure}{0.9\textwidth} 
        \centering
        \includegraphics[width=\textwidth]{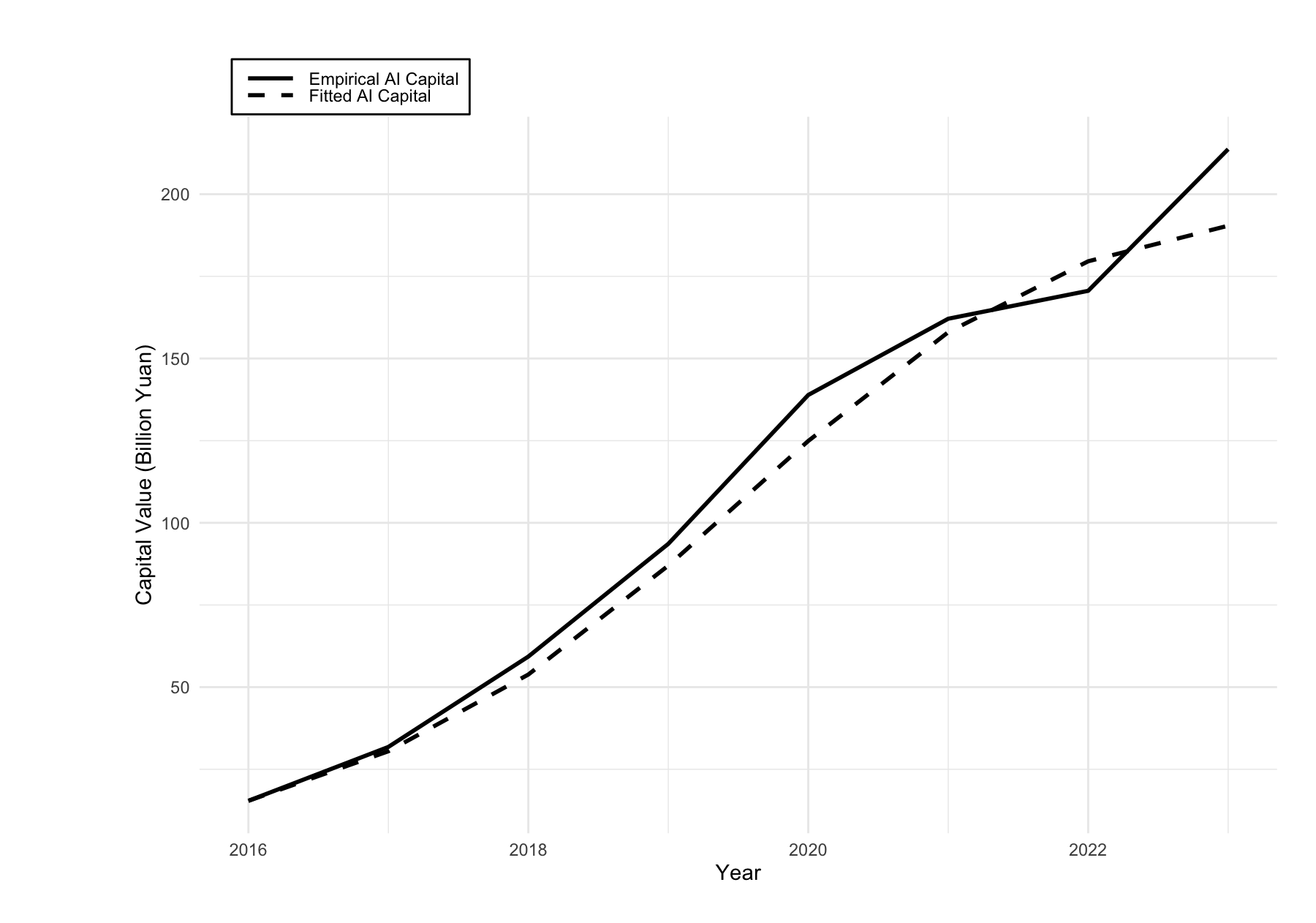} 
        \caption{Empirical versus fitted AI capital.}
        \label{fig1}
    \end{subfigure}
    \hfill
    \begin{subfigure}{0.9\textwidth} 
        \centering
        \includegraphics[width=\textwidth]{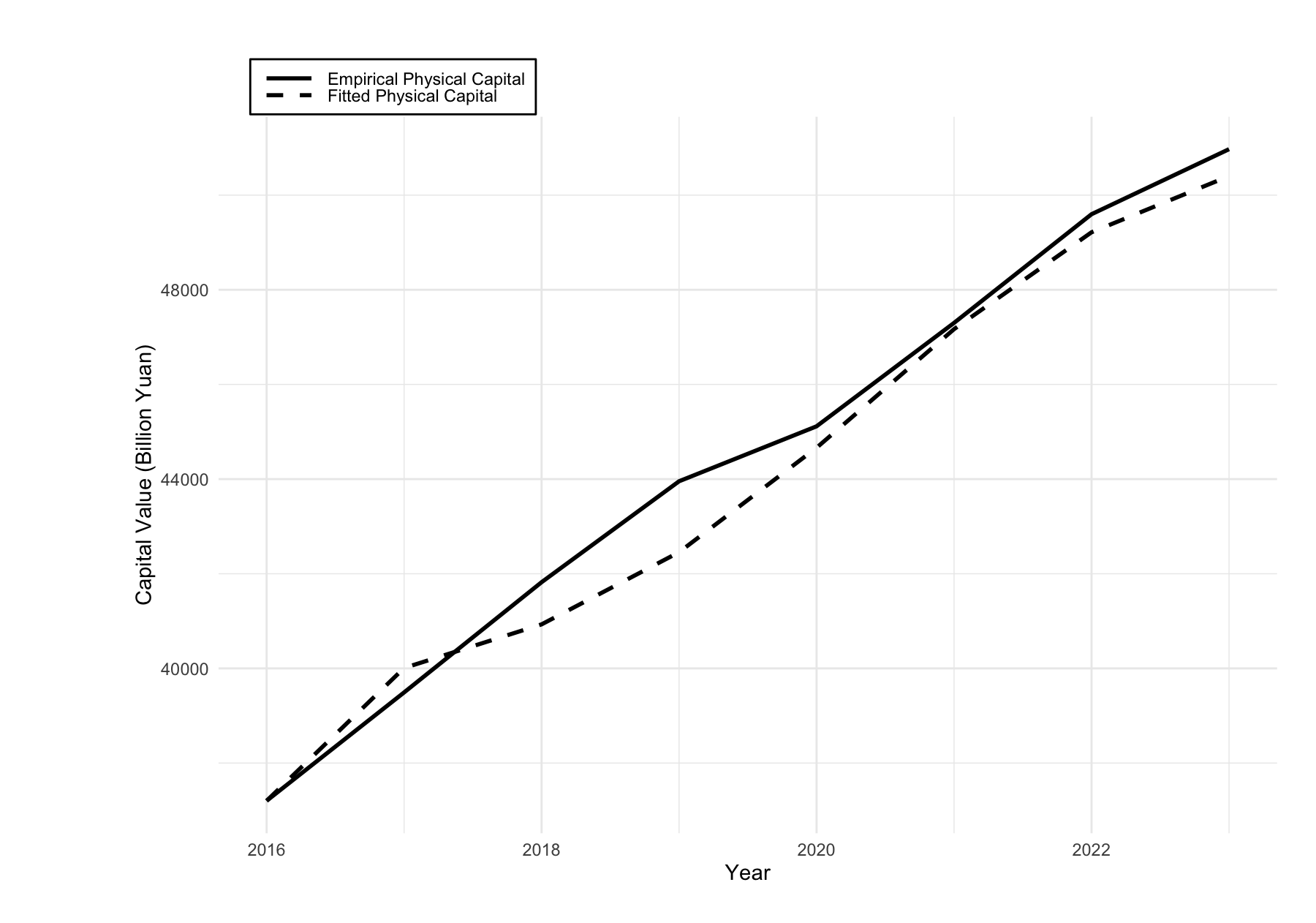} 
        \caption{Empirical versus fitted physical capital.}
        \label{fig2}
    \end{subfigure}
    \caption{Comparison of empirical and fitted AI capital and physical capital.}
    \label{fig:main1}
\end{figure}

\begin{figure}[htbp]
    \centering %
    \begin{subfigure}{0.9\textwidth} 
        \centering
        \includegraphics[width=\textwidth]{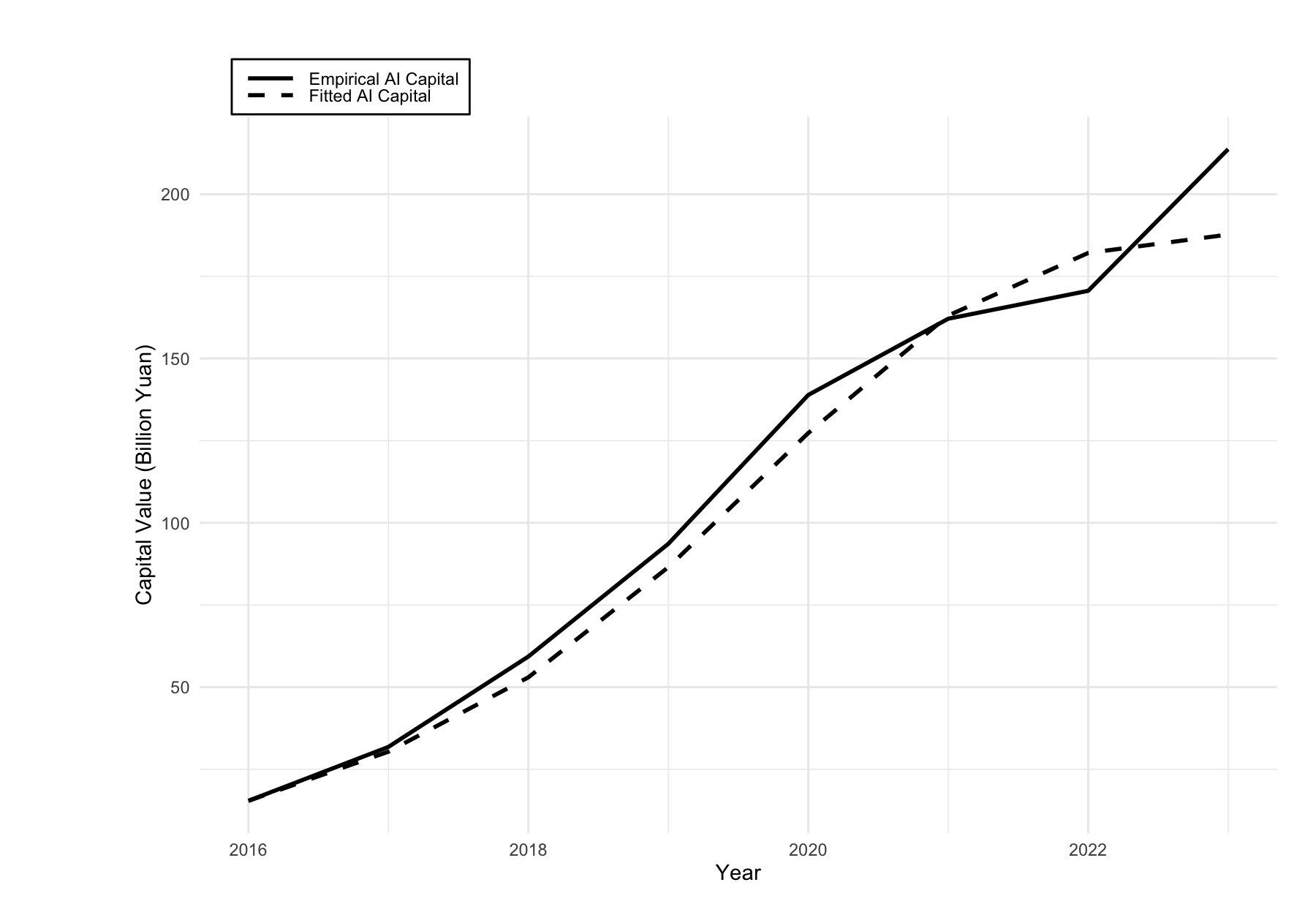} 
        \caption{Empirical versus fitted AI capital.}
        \label{fig3}
    \end{subfigure}
    \hfill
    \begin{subfigure}{0.9\textwidth}
        \centering
        \includegraphics[width=\textwidth]{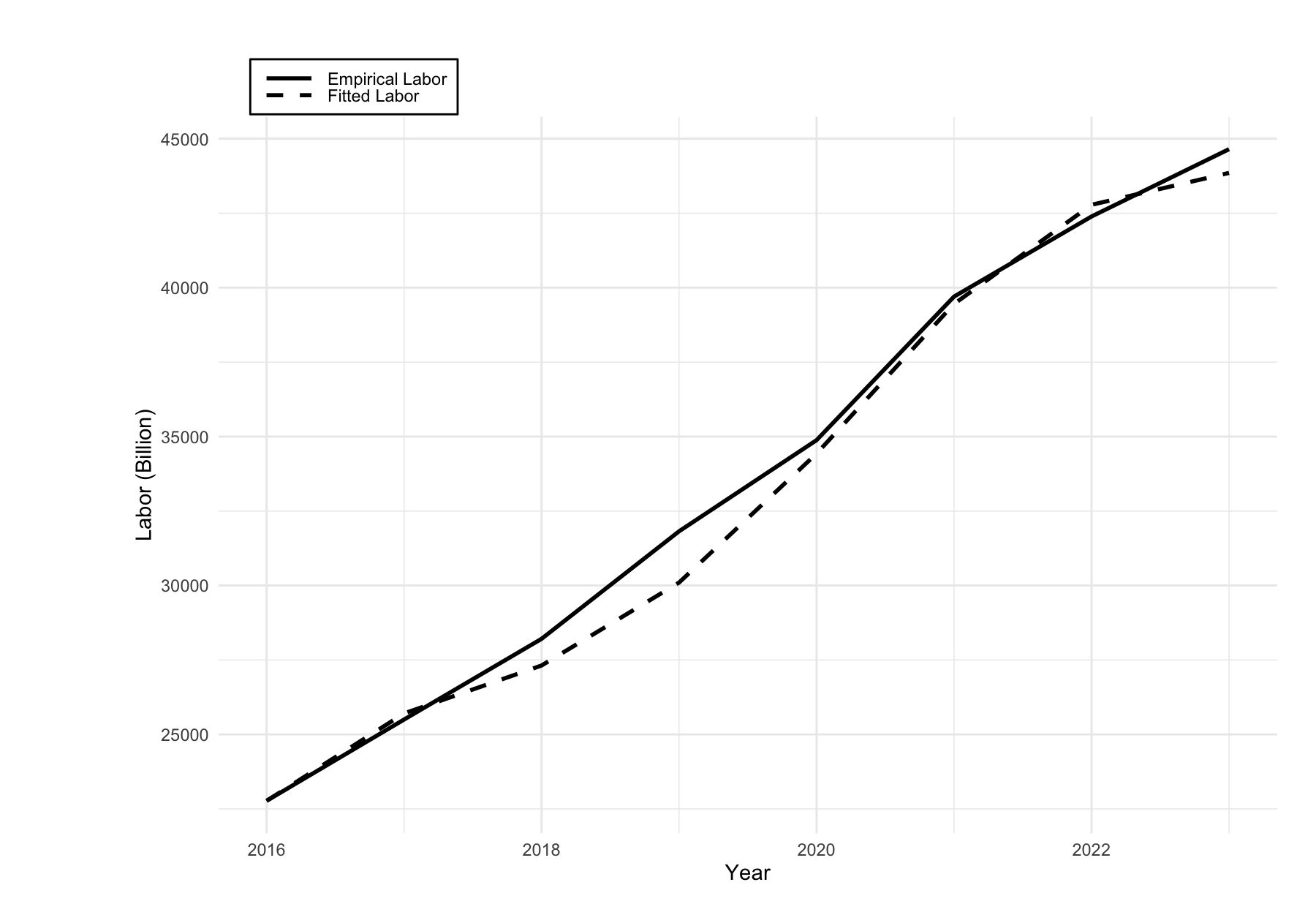} 
        \caption{Empirical versus fitted physical capital.}
        \label{fig4}
    \end{subfigure}
    \caption{Comparison of empirical and fitted AI capital and labor.}
    \label{fig:main2}
\end{figure}

The fitted values for AI capital, physical capital, and labor are derived from the discrete Lotka-Volterra equations \eqref{discrete1} and \eqref{discrete2}, using the estimated coefficients presented in Table \ref{table3}. These results are visualized in Figures \ref{fig:main1} and \ref{fig:main2}, demonstrating the model’s alignment with empirical trends.

To quantitatively assess the model’s accuracy, we employ the Mean Absolute Percentage Error (MAPE) metric, defined as
$$ \mbox{MAPE}=\frac{1}{n} \sum_{i=1}^n\left|\frac{w_i-\hat{w}_i}{w_i}\right| \times 100 \%,$$
where $w_i$ represents the observed empirical values and  $\hat{w}_i$ denotes the model’s predicted values. MAPE measures the average percentage deviation between the model and data, with smaller values indicating superior predictive performance. As shown in Table \ref{table4}, the low MAPE values across all variables confirm that the model robustly captures the underlying dynamics of AI capital, physical capital, and labor interactions.
 \begin{table}[htbp]
 \centering
 \begin{tabular}{|c|c|c| }
 \hline 
 & AI Capital & Physical Capital  \\
\hline 
MAPE   & $6.15\%$  &  $1.25\%$ \\
 \hline  
  & AI Capital & Labor  \\
\hline 
MAPE   & $ 6.33\%$  &  $ 1.75\%$ \\
 \hline  
 \end{tabular}
 \caption{MAPE values for empirical and fitted values in two models. }
 \label{table4}
 \end{table}

\subsection{AI capital and physical capital}
\label{subsection1}
First, the interaction between AI capital and physical capital is explored. We let $x$ represent AI capital and $y$ stand for physical capital. With the above obtained regression coefficients, we arrive at the following system of Lotka-Volterra equations 
\begin{align}
& \frac{d x}{d t}=3.852613 x(t)-0.006965 x(t)^2-0.000048 x(t) y(t),  \label{fteqn1} \\
& \frac{d y}{d t}=4.934909 y(t)-0.000126 y(t)^2+0.007846 y(t)x(t).
\label{fteqn2}
\end{align}
The signs of the coefficients $b_{12}$ and $b_{21}$ in equations \eqref{fteqn1} and \eqref{fteqn2} reveal a predator-prey dynamic between AI capital and physical capital in the long run.  In more details, $b_{12}=-0.000048$ implies that physical capital negatively impacts AI capital growth, analogous to a predator depleting its prey, while $b_{21}=0.007846$  positively stimulates physical capital accumulation, similar with how prey sustains its predator. The small magnitude $b_{12}$ relative to $b_{21}$ further implies that the predatory effect of physical capital is weak compared to its dependency on AI-driven stimulation. These results confirm to the reality of China that the benefits of AI and automation indeed promotes the development of the traditional industries in China such as in electric vehicle, smart phone manufacturing, \emph{etc}.  On the other hand, because traditional industries can sustain more employment, the national policy prohibits local authorities from providing indiscriminate subsidies to purely AI-focused projects and  prevents tech giants from monopolizing resources, which allows traditional industries to maintain a strong competitive position against AI industries. The results can be further verified in the follow-up Subsection \ref{subsection2}.

\subsection{AI capital and labor}
\label{subsection2}
Next, we analyze the dynamics of AI capital  and labor. We denote AI capital by $x$ and labor by $y$. Using empirically estimated coefficients, the system is governed by 
\begin{align}
& \frac{d x}{d t}=3.741844 x(t)-0.000943 x(t)^2-0.000081 x(t) y(t),  \label{fteqn3} \\
& \frac{d y}{d t}=4.480796 y(t)-0.000187 y(t)^2+0.020083 y(t)x(t).
\label{fteqn4}
\end{align}
It follows from $b_{12}<0$ and $b_{21}>0$ that AI capital and labor interacts in a predator-prey manner in the long term. Specifically, $b_{21}=0.020083$ indicates that AI capital contributes to wage growth, while $b_{12}=-0.000008$ suggests that labor exerts a weak predatory effect on AI capital. These findings match with the above results in Subsection \ref{subsection1}. Our model shows that AI capital in China can raise the overall wage level, not surprisingly, aligning with numerous existing arguments \cite{jones, ai1, bryn2021, bughin2018} suggesting that AI enhances total factor productivity. However, our model also reveals that labor weakly inhibits AI capital. This is mainly due to that, as discussed in Subsection \ref{subsection1}, the national policy aligns AI innovation with labor market stabilization and steers AI governance toward employment-friendly outcomes.



\section{Equilibrium analysis}
\label{s4}
In this section, we analyze the equilibrium points of the two systems, namely, equations \eqref{fteqn1}-\eqref{fteqn2}, and equations \eqref{fteqn3}-\eqref{fteqn4}. The Lotka-Volterra system in equations \eqref{lveqn1}-\eqref{lveqn2} will reach the equilibrium in the long run when the size of species becomes constant, namely, we arrive at the following system of equations
\begin{align}
x(t)(a_1+b_{11}x(t)+b_{12}y(t))=0, \label{nullline1} \\
y(t)(a_2+b_{21}x(t)+b_{22}y(t))=0. \label{nullline2}
\end{align}
Excluding the invariant lines $x=0$ and $y=0$, the non-trivial equilibrium point of our interest is 
\begin{equation}
(x^*,y^*)=\left( \frac{a_1b_{22}-b_{12}a_2}{b_{12}b_{21}-b_{11}b_{22}}, \frac{b_{11}a_2-a_1b_{21}}{b_{12}b_{21}-b_{11}b_{22}} \right).
\label{equil13}
\end{equation}
It follows from the Hartman-Grobman theorem \cite{smale2013} that a nonlinear system of ordinary differential equations (ODEs) can be locally approximated by its linearization at an equilibrium point. The Jacobian matrix  of the linearized system allows us to qualitatively assess the stability of an equilibrium. The Jacobian matrix of the system in equations \eqref{lveqn1} and \eqref{lveqn2} is 
\begin{equation}
J(x,y)=\begin{bmatrix} 
a_1+2b_{11}x+b_{12}y & b_{12}x \\
b_{21}y & a_2+b_{21}x + 2b_{22}y \\
\end{bmatrix}.
\end{equation}

\subsection{AI capital and physical capital}
\begin{figure}[htbp]
\centering
\includegraphics[width=0.85\textwidth]{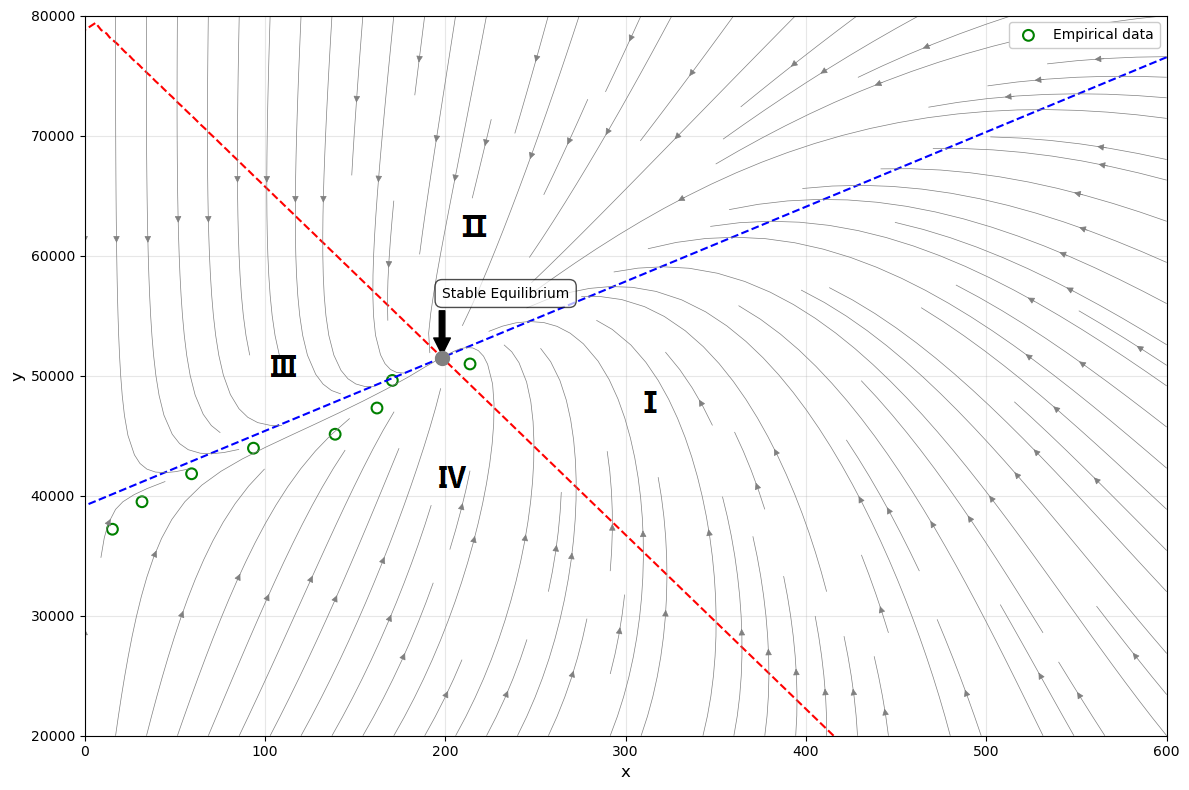}
\caption{The evolution trend of AI capital and physical capital in the long run.}
\label{phase1}
\end{figure}

The specific dynamics of AI capital ($x$) and physical capital ($y$) are presented in equations \eqref{fteqn1} and \eqref{fteqn2}. The corresponding nontrivial equilibrium point is determined by the following null lines 
\begin{align}
& 3.852613-0.006965 x-0.000048 y=0,  \label{nullline3} \\
& 4.934909-0.000126 y+0.007846 x=0. \label{nullline4}
\end{align} 
and we find the equilibrium point $(x_1^*,y_1^*)=(198.18, 51506.42)$. The corresponding eigenvalues  $\lambda_1=-2.29$ and $\lambda_2=-5.57$ indicate that $(x_1^*,y_1^*)$ is a stable node.  As shown in Figure \ref{phase1}, the null lines \eqref{nullline3} and \eqref{nullline4} are represented as the red and blue dashed lines. It can be seen that the null lines partition the first quadrant into 4 regions. In Region I, $\frac{dx}{dt}<0$ and $\frac{dy}{dt}>0$, showing that AI capital decreases and physiscal capital increases. We can find  in Region II that both AI capital and physical capital decrease   since $\frac{dx}{dt}<0$ and $\frac{dy}{dt}<0$. It follows from $\frac{dx}{dt}>0$ and $\frac{dy}{dt}<0$ that AI capital increases and physical capital decreases in Region III. Both AI capital and physical capital increase in Region IV since $\frac{dx}{dt}>0$ and $\frac{dy}{dt}>0$. The model demonstrates that AI capital and physical capital in China interact in a predator-prey manner. The identification of a stable node, rather than cycles or unstable equilibria, suggests that the current policy mix is effective in achieving a predictable convergence path. This stability, however, may come at the cost of structural rigidity, limiting the system’s flexibility to adapt to optimal resource allocations, a key consideration for long-term policy design. 


\subsection{AI capital and labor}
\begin{figure}[htbp]
\centering
\includegraphics[width=0.85\textwidth]{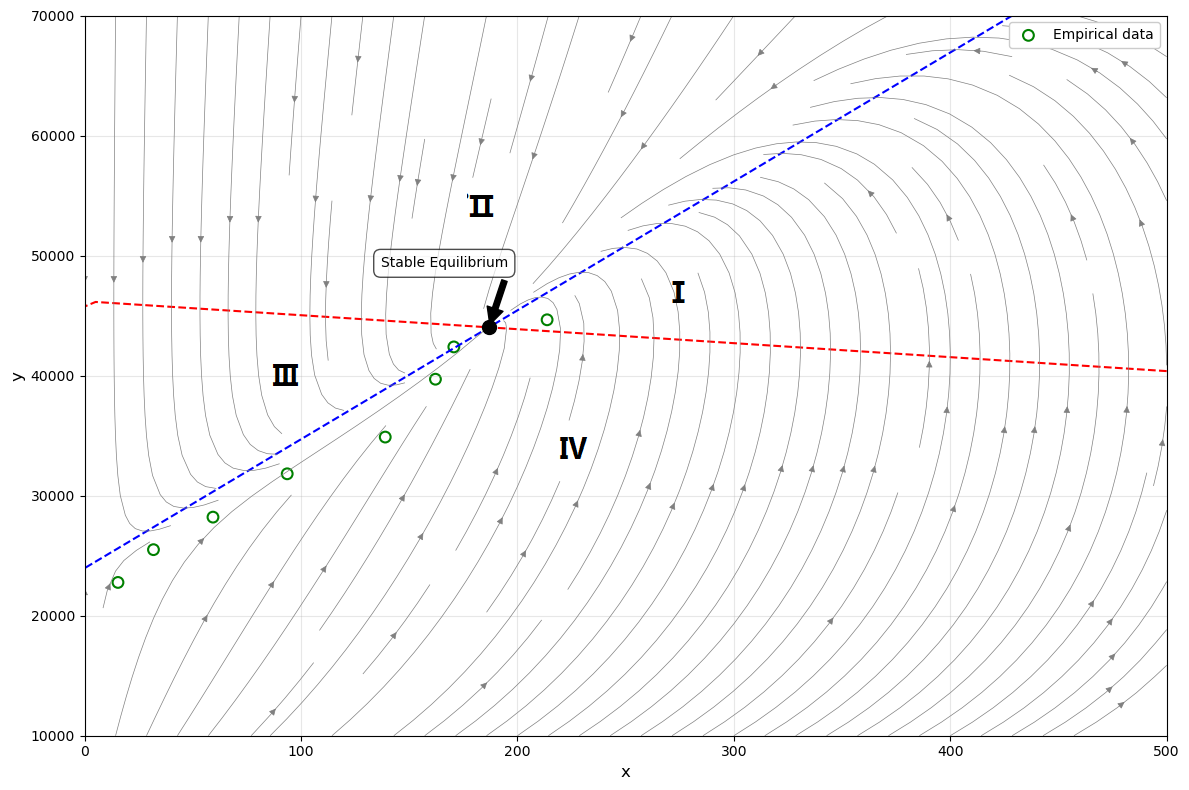}
\caption{The evolution trend of AI capital and labor in the long run.}
\label{phase2}
\end{figure}

The interaction between AI capital ($x$) and labor ($y$) is presented in equations \eqref{fteqn3} and \eqref{fteqn4}. The corresponding null lines are 
\begin{align}
&  3.741844 -0.000943 x(t)-0.000081  y(t)=0,  \label{nullline5} \\
& 4.480796 -0.000187 y(t)+0.020083 x(t)=0.
\label{nullline6}
\end{align}
The equilibrium point is $(x^*_2,y^*_2)=(186.78, 44021.09)$ and the eigenvalues are $\lambda_1=-2.52$ and $\lambda_2=-5.89$. The results show that $(x_2^*,y_2^*)$ is a stable node. The null lines \eqref{nullline5} and \eqref{nullline6} are presented as the red and blue dashed lines in Figure \ref{phase2}. We can notice that in Region I, since \(\frac{dx}{dt} < 0\) and \(\frac{dy}{dt} > 0\), AI capital declines while labor expands.  Observing Region II, where \(\frac{dx}{dt} < 0\) and \(\frac{dy}{dt} < 0\), we see that both AI capital and labor decrease.  Because \(\frac{dx}{dt} > 0\) and \(\frac{dy}{dt} < 0\), AI capital grows while labor contracts in Region III. An increase in both AI capital and labor occurs in Region IV, as indicated by \(\frac{dx}{dt} > 0\) and \(\frac{dy}{dt} > 0\). Our model indicates that the relationship between AI capital and labor in China follows a predator-prey dynamic. The empirical data shown in Figure \ref{phase2} support our results. The results demonstrate that the equilibrium state is predominantly shaped by state intervention. While government policies effectively stabilize labor market dynamics, they simultaneously impose constraints on the development of AI capital. This dual effect creates a trade-off between market stability and technological growth potential. 

\section{Global Sensitivity Methodology}
\label{s5}
To inform robust policy design, we evaluate the stability of our findings and identify key policy leverage points through a Sobol global sensitivity analysis \cite{salte2010}. This method quantifies how uncertainty in estimated interaction parameters translates into uncertainty in long-run equilibrium outcomes, highlighting which relationships policymakers should monitor most closely.


We notice from the equation \eqref{equil13} that the equilibrium states are determined by the six parameters in the LV equations 
$$ \mathbf{Y}=f(\boldsymbol{\theta})=  \begin{bmatrix}
x^* \\
y^*
\end{bmatrix},$$
where $ \boldsymbol{\theta}=(a_1, b_{11}, b_{12}, a_2, b_{21}, b_{22})$.  This Sobol method decomposes the total variance  of the equilibrium output $V(\mathbf{Y})$  into additive and interaction terms associated with each model parameter $\theta_i$, $i=1,\ldots,6$. The purpose of this test is to quantify the relative influence of each model parameter $(a_1, b_{11}, b_{12}, a_2, b_{21}, b_{22})$ on the steady-state equilibrium of AI capital $x^*$ and its counterpart variable $y^*$ (physical capital or labor). This approach provides a systematic framework to assess how uncertainty in the parameters propagates through the nonlinear system and affects equilibrium outcomes.

The first-order and total-order sensitivity indices are respectively given by
\begin{equation}
S_i = \frac{V_i}{V(\mathbf{Y})}, \qquad S_{Ti} = 1 - \frac{V_{\sim i}}{V(\mathbf{Y})},
\end{equation}
where $V_i$ denotes the partial variance caused by parameter $i$, $V_{\sim i}$ is the variance of the output when $i$ is fixed, and $V(\mathbf{Y})$ represents the total variance of the equilibrium vector $\mathbf{Y}= [x^* \ y^*]^T$.

The analysis was implemented using the Saltelli sampling algorithm \cite{salte2010} in Python SALib, with a base sample size of ($N = 1024$), yielding $6(N+2)$ evaluations for each subsystem, AI–capital and AI–labor.
The six parameters $(a_1, b_{11}, b_{12}, a_2, b_{21}, b_{22})$ were uniformly perturbed within $\pm 10\%$ around their estimated baseline values (see Table \ref{ci}, which lists all coefficients and confidence intervals).

\begin{sidewaystable}[htbp]
\centering
\label{tab:params}
\begin{tabular}{lccc}
\hline
\textbf{Parameter} & \textbf{Symbol} & \textbf{Estimated Value} & \textbf{95\,\%\,CI} \\
\hline
\multicolumn{4}{l}{\textit{(a) AI--Physical Capital Subsystem}}\\
Intrinsic growth rate (AI capital) & $a_1$  & 3.8526 & [3.35,\;4.35] \\
Self-limiting coefficient (AI capital) & $b_{11}$ & $-6.97\times10^{-3}$ & [$-8.36\times10^{-3}$,\;$-5.58\times10^{-3}$] \\
Interaction coefficient (AI $\rightarrow$ capital) & $b_{12}$ & $-4.8\times10^{-5}$ & [$-5.8\times10^{-5}$,\;$-3.8\times10^{-5}$] \\
Intrinsic growth rate (physical capital) & $a_2$  & 4.9349 & [4.45,\;5.42] \\
Interaction coefficient (capital $\rightarrow$ AI) & $b_{21}$ & $7.85\times10^{-3}$ & [$6.30\times10^{-3}$,\;$9.40\times10^{-3}$] \\
Self-limiting coefficient (physical capital) & $b_{22}$ & $-1.26\times10^{-4}$ & [$-1.51\times10^{-4}$,\;$-1.01\times10^{-4}$] \\
\hline
\multicolumn{4}{l}{\textit{(b) AI--Labor Subsystem}}\\
Intrinsic growth rate (AI capital) & $a_1$  & 3.7418 & [3.20,\;4.28] \\
Self-limiting coefficient (AI capital) & $b_{11}$ & $-9.43\times10^{-4}$ & [$-1.13\times10^{-3}$,\;$-7.54\times10^{-4}$] \\
Interaction coefficient (AI $\rightarrow$ labor) & $b_{12}$ & $-8.1\times10^{-5}$ & [$-9.7\times10^{-5}$,\;$-6.5\times10^{-5}$] \\
Intrinsic growth rate (labor) & $a_2$  & 4.4808 & [4.00,\;4.96] \\
Interaction coefficient (labor $\rightarrow$ AI) & $b_{21}$ & $2.01\times10^{-2}$ & [$1.61\times10^{-2}$,\;$2.41\times10^{-2}$] \\
Self-limiting coefficient (labor) & $b_{22}$ & $-1.87\times10^{-4}$ & [$-2.24\times10^{-4}$,\;$-1.50\times10^{-4}$] \\
\hline
\end{tabular}
\caption{Baseline parameter estimates and 95\,\% confidence intervals for the AI--capital and AI--labor Lotka--Volterra subsystems (2016--2023).}
\label{ci}
\end{sidewaystable}

\subsection{The AI-physical capital subsystem}
For each sampled parameter set, the equilibrium point $(x^*, y^*)$ was computed analytically from the steady-state LV equations \eqref{equil13}. Samples yielding non-finite or negative equilibria were automatically rejected to preserve physical consistency.

 For each valid realization, the resulting equilibria were recorded and used to compute the partial and total variances.
  For example, when $a_1$ was perturbed by $+5 \%$, that is, $(a_1' = 1.05a_1)$, the model produced $x^* = 205.1$ and $y^* = 52 084$; when $b_{21}$ was decreased by $10\%$ ($b_{21}' = 0.9b_{21}$), the equilibrium shifted to $x^* = 191.3$, $y^* = 50 762$.
The variance across all such realizations defines $V_i$ and $V_{\sim i}$ for the corresponding parameter.

\begin{table}[htbp]
\centering
\renewcommand\arraystretch{1.1}
\setlength{\tabcolsep}{5pt}
\begin{tabular}{lccc}
\hline
\textbf{Parameter} & \textbf{Symbol} & \textbf{First--Order $S_i$} & \textbf{Total--Order $S_{Ti}$} \\
\hline
\makecell[l]{Intrinsic growth rate \\ (AI capital)} & $a_1$ & 0.328 & 0.334 \\
\makecell[l]{Self-limiting coefficient \\ (AI capital)} & $b_{11}$ & 0.079 & 0.085 \\
\makecell[l]{Interaction coefficient \\ (AI $\rightarrow$ capital)} & $b_{12}$ & 0.245 & 0.249 \\
\makecell[l]{Intrinsic growth rate \\ (physical capital)} & $a_2$ & 0.080 & 0.083 \\
\makecell[l]{Interaction coefficient \\ (capital $\rightarrow$ AI)} & $b_{21}$ & 0.008 & 0.008 \\
\makecell[l]{Self-limiting coefficient \\ (physical capital)} & $b_{22}$ & 0.250 & 0.254 \\
\hline
\multicolumn{2}{l}{\textbf{Sum of $S_i$}} & \multicolumn{2}{c}{0.990} \\
\hline
\end{tabular}
\caption{First-- and total--order Sobol sensitivity indices ($S_i$, $S_{Ti}$) for the equilibrium variable $x^*$ in the AI--physical capital subsystem.}
\label{tab:sobol_x_phys}
\end{table}

\begin{table}[htbp]
\centering
\renewcommand\arraystretch{1.1}
\setlength{\tabcolsep}{5pt}
\begin{tabular}{lccc}
\hline
\textbf{Parameter} & \textbf{Symbol} & \textbf{First--Order $S_i$} & \textbf{Total--Order $S_{Ti}$} \\
\hline
\makecell[l]{Intrinsic growth rate \\ (AI capital)} & $a_1$ & 0.134 & 0.137 \\
\makecell[l]{Self-limiting coefficient \\ (AI capital)} & $b_{11}$ & 0.031 & 0.033 \\
\makecell[l]{Interaction coefficient \\ (AI $\rightarrow$ capital)} & $b_{12}$ & 0.103 & 0.106 \\
\makecell[l]{Intrinsic growth rate \\ (physical capital)} & $a_2$ & 0.174 & 0.175 \\
\makecell[l]{Interaction coefficient \\ (capital $\rightarrow$ AI)} & $b_{21}$ & 0.016 & 0.018 \\
\makecell[l]{Self-limiting coefficient \\ (physical capital)} & $b_{22}$ & 0.534 & 0.542 \\
\hline
\multicolumn{2}{l}{\textbf{Sum of $S_i$}} & \multicolumn{2}{c}{0.992} \\
\hline
\end{tabular}
\caption{First-- and total--order Sobol sensitivity indices ($S_i$, $S_{Ti}$) for the equilibrium variable $y^*$ in the AI--physical capital subsystem.}
\label{tab:sobol_y_phys}
\end{table}

For the AI equilibrium $x^*$,  the variance is primarily driven by the intrinsic growth of AI capital ($a_1, S_{T i}=0.334$) and the self-limiting effect of physical capital ($b_{22}, S_{T i}=0.254$).
The AI $\rightarrow$ physical capital interaction term ( $b_{12}$ ) also contributes substantially with $S_{T i}=0.249$, indicating that both Al's internal expansion dynamics and its spillover to the physical sector jointly govern the equilibrium AI level.
Other parameters, including the feedback interaction $b_{21}$ and capital's own growth rate ($a_2$), exert smaller influences ($S_{T i}<0.09$ ).

In contrast, the variance pattern of the physical capital equilibrium $y^*$ becomes more concentrated. The total order variance $S_{T i}=0.542$ shows that the self-limiting coefficient of physical capital  $b_{22}$dominates, followed by capital's intrinsic growth $a_2$ ($  S_{T i}=0.175$ ) and the AI$\rightarrow$ capital coupling $b_{12}$ ($S_{T i}=0.106$). 

Also, we note that parameters associated with Al itself (such as $a_1, b_{11}$) have secondary effects.

\begin{figure}[htbp]
\centering
\subfloat[\textbf{$x^*$:} Sobol sensitivity indices for the equilibrium AI variable in the AI--physical capital subsystem.]
{\includegraphics[width=0.8\textwidth]{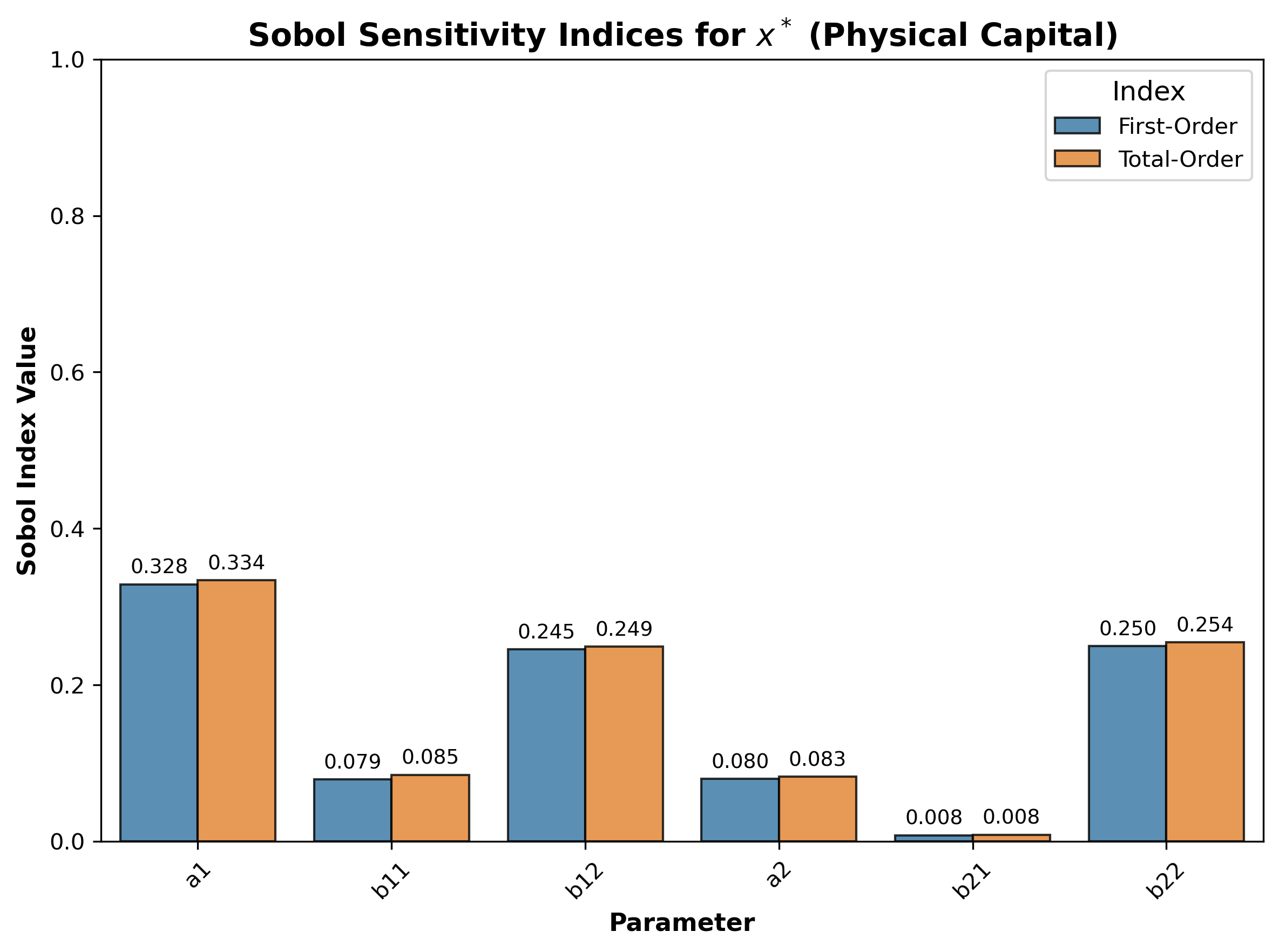}\label{fig:sobol_x_physical}}\\[1em]
\subfloat[\textbf{$y^*$:} Sobol sensitivity indices for the equilibrium physical capital variable in the AI--physical capital subsystem.]
{\includegraphics[width=0.8\textwidth]{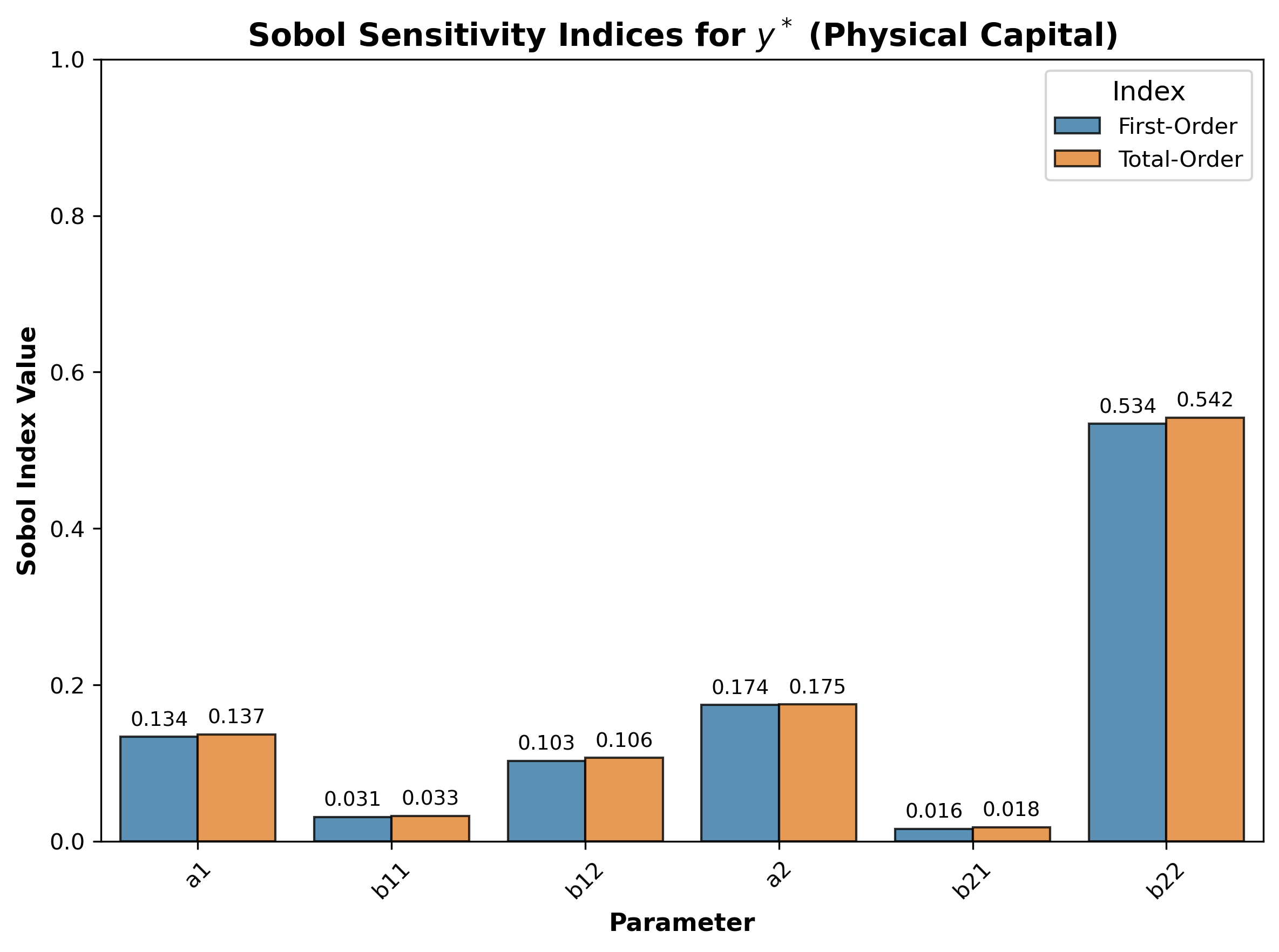}\label{fig:sobol_y_physical}}
\caption{First-- and total--order Sobol sensitivity indices ($S_i$, $S_{Ti}$) for the AI--physical capital subsystem.}
\label{sobol_physical}
\end{figure}

This asymmetry indicates that equilibrium physical capital output is mainly regulated by its internal saturation and growth feedback rather than by Al's parameters, underscoring a hierarchical structure in which AI indirectly influences physical capital through coupling terms. The corresponding Sobol indices for both $x^*$ and $y^*$ are visualized in Fig. \ref{sobol_physical}.

\subsection{The AI--labor subsystem}
A parallel Sobol analysis was conducted for the AI--labor subsystem, in which $x^*$ represents the equilibrium level of AI capital and $y^*$ denotes the equilibrium level of labor.

\begin{table}[htbp]
\centering
\renewcommand\arraystretch{1.1}
\setlength{\tabcolsep}{5pt}
\begin{tabular}{lccc}
\hline
\textbf{Parameter} & \textbf{Symbol} & \textbf{First--Order $S_i$} & \textbf{Total--Order $S_{Ti}$} \\
\hline
\makecell[l]{Intrinsic growth rate \\ (AI capital)} & $a_1$ & 0.208 & 0.216 \\
\makecell[l]{Self-limiting coefficient \\ (AI capital)} & $b_{11}$ & 0.002 & 0.001 \\
\makecell[l]{Interaction coefficient \\ (AI $\rightarrow$ labor)} & $b_{12}$ & 0.347 & 0.354 \\
\makecell[l]{Intrinsic growth rate \\ (labor)} & $a_2$ & 0.056 & 0.056 \\
\makecell[l]{Interaction coefficient \\ (labor $\rightarrow$ AI)} & $b_{21}$ & 0.045 & 0.045 \\
\makecell[l]{Self-limiting coefficient \\ (labor)} & $b_{22}$ & 0.333 & 0.344 \\
\hline
\multicolumn{2}{l}{\textbf{Sum of $S_i$}} & \multicolumn{2}{c}{0.989} \\
\hline
\end{tabular}
\caption{First-- and total--order Sobol sensitivity indices ($S_i$, $S_{Ti}$) for the equilibrium variable $x^*$ in the AI--labor subsystem.}
\label{tab:sobol_x_labor}
\end{table}

\begin{table}[htbp]
\centering
\renewcommand\arraystretch{1.1}
\setlength{\tabcolsep}{5pt}
\begin{tabular}{lccc}
\hline
\textbf{Parameter} & \textbf{Symbol} & \textbf{First--Order $S_i$} & \textbf{Total--Order $S_{Ti}$} \\
\hline
\makecell[l]{Intrinsic growth rate \\ (AI capital)} & $a_1$ & 0.368 & 0.375 \\
\makecell[l]{Self-limiting coefficient \\ (AI capital)} & $b_{11}$ & 0.002 & 0.002 \\
\makecell[l]{Interaction coefficient \\ (AI $\rightarrow$ labor)} & $b_{12}$ & 0.614 & 0.619 \\
\makecell[l]{Intrinsic growth rate \\ (labor)} & $a_2$ & 0.001 & 0.001 \\
\makecell[l]{Interaction coefficient \\ (labor $\rightarrow$ AI)} & $b_{21}$ & 0.000 & 0.001 \\
\makecell[l]{Self-limiting coefficient \\ (labor)} & $b_{22}$ & 0.007 & 0.008 \\
\hline
\multicolumn{2}{l}{\textbf{Sum of $S_i$}} & \multicolumn{2}{c}{0.992} \\
\hline
\end{tabular}
\caption{First-- and total--order Sobol sensitivity indices ($S_i$, $S_{Ti}$) for the equilibrium variable $y^*$ in the AI--labor subsystem.}
\label{tab:sobol_y_labor}
\end{table}

The AI-labor subsystem demonstrates significantly tighter coupling compared to the AI-physical capital subsystem. In both of its equilibria, the system is dominated by the $\mathrm{Al} \rightarrow$ labor interaction parameter ($b_{12}$) and the intrinsic AI growth rate ($a_1$). 

A detailed analysis of the $x^*$ equilibrium reveals that three parameters account for the majority of the variance: the interaction term $b_{12}$ ($S_{T i}=0.354$), the labor self-limiting coefficient $b_{22}\left(S_{T i}=0.344\right)$, and the AI growth rate $a_1$ ( $S_{T i}=0.216$ ). This result highlights a feedback mechanism in which the expansion of Al is regulated not only by its own growth potential but also by the adaptive capacity and saturation dynamics of the labor force.

For $y^*$, the dominance is stark: $b_{12}$ and $a_1$ account for $61 \%\left(S_{T_i}=0.619\right)$ and $37 \%\left(S_{T_i}=0.375\right)$ of the total variance, respectively. The remaining parameters have a negligible effect.

Thus, the labor equilibrium is governed almost exclusively by AI-driven interactions. This implies that the steady-state level of labor is determined predominantly by the intensity of technological inputs and AI productivity, while labor's intrinsic growth and saturation parameters have a negligible influence.

\subsection{Cross-subsystem analysis}
The results illustrate a definitive structural shift from a balanced, partially coupled regime in the Al-physical capital system-where equilibrium is shaped by both Al cross-effects and capital's self-limiting feedback-to a unidirectional, technology-dominant regime in the AI-labor system, where variance is overwhelmingly concentrated in AI-related parameters.

\begin{figure}[htbp]
\centering
\subfloat[\textbf{$x^*$:} Sobol sensitivity indices for the equilibrium AI variable in the AI--labor subsystem.]
{\includegraphics[width=0.8\textwidth]{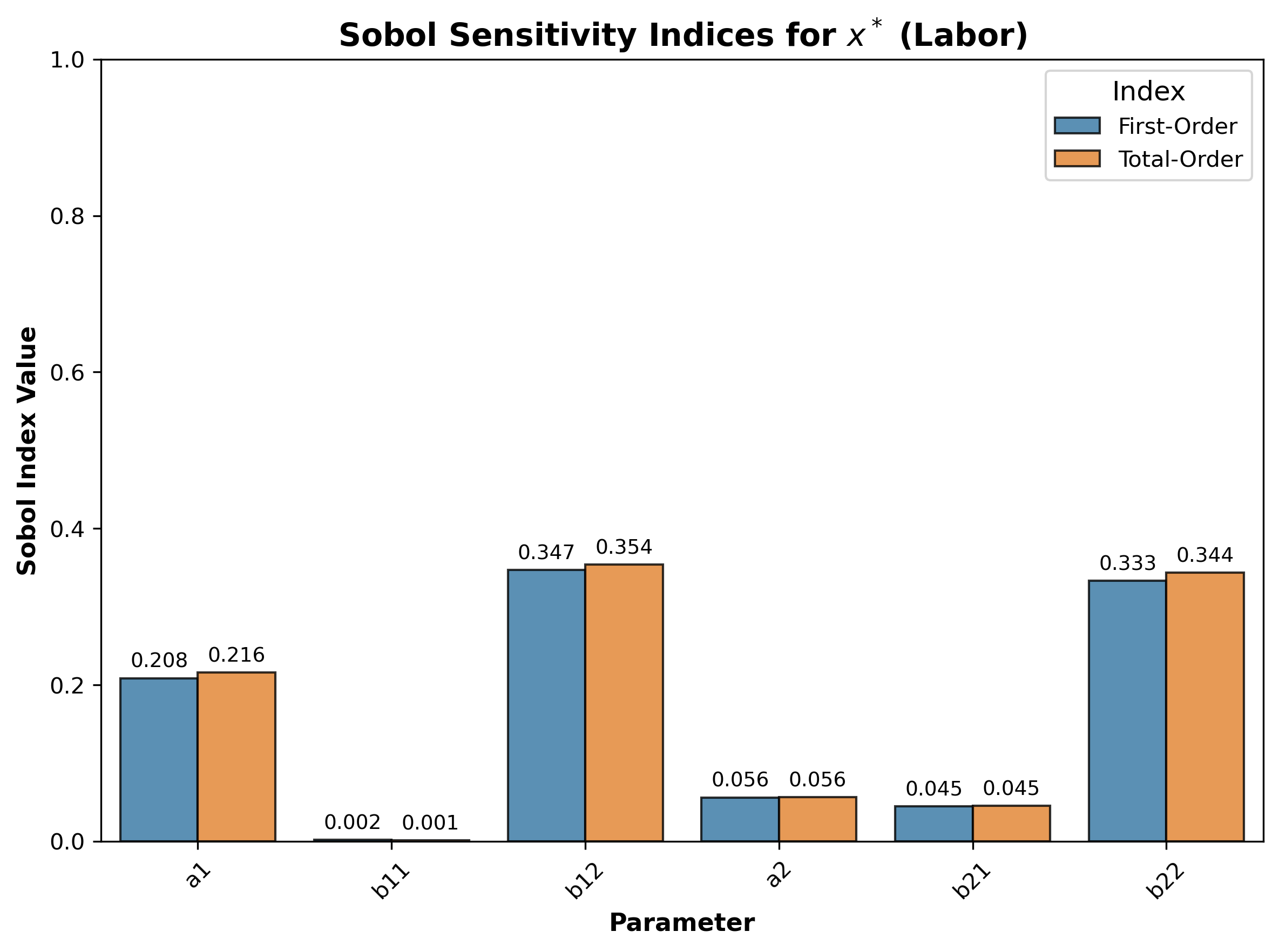}\label{fig:sobol_x_labor}}\\[1em]
\subfloat[\textbf{$y^*$:} Sobol sensitivity indices for the equilibrium labor variable in the AI--labor subsystem.]
{\includegraphics[width=0.8\textwidth]{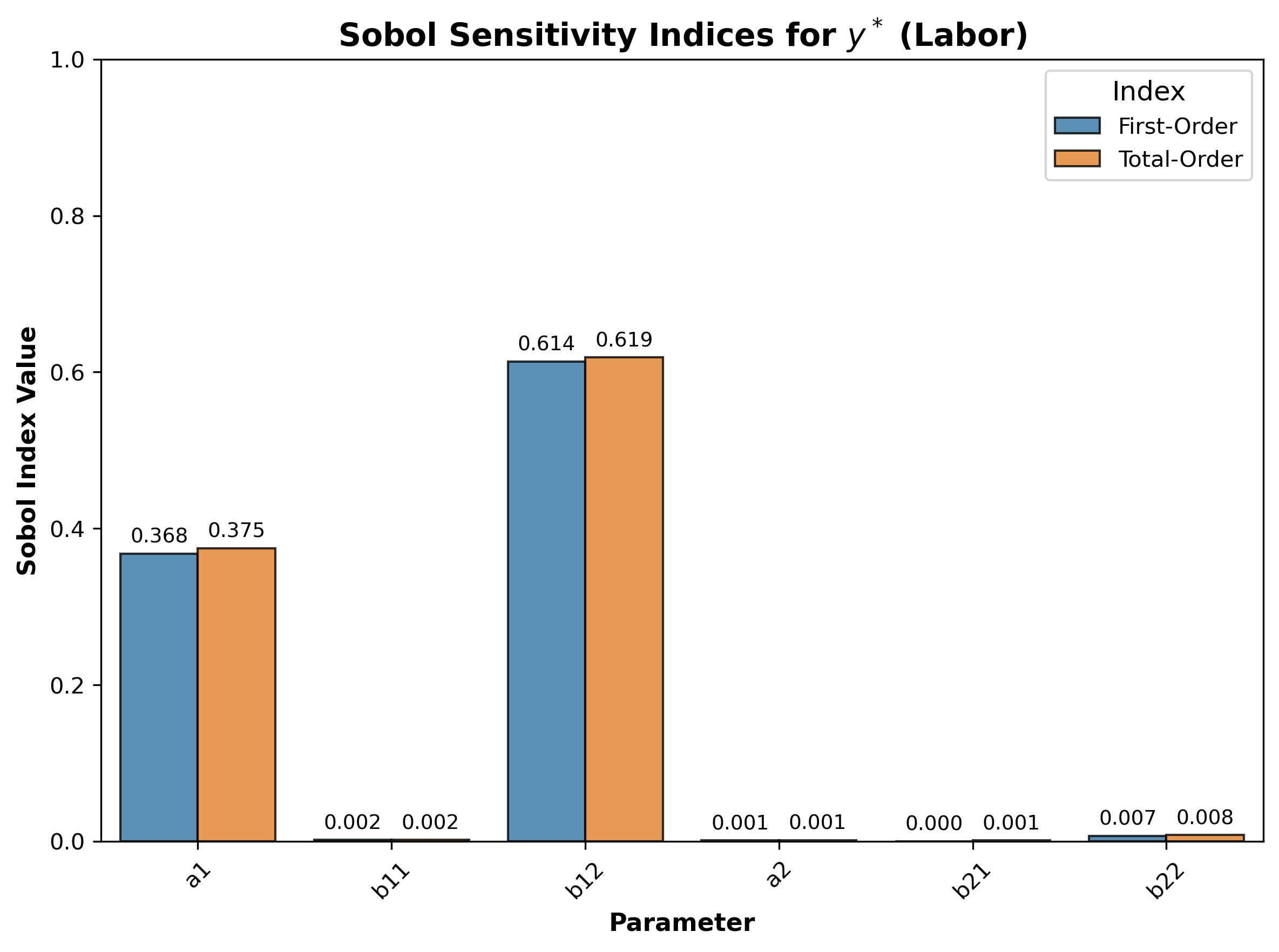}\label{fig:sobol_y_labor}}
\caption{First-- and total--order Sobol sensitivity indices ($S_i$, $S_{Ti}$) for the AI--labor subsystem.}
\label{sobol_labor}
\end{figure}

This transition is quantitatively demonstrated by the evolving role of the AI-driven interaction term $b_{12}$, which shifts from a moderate influence ($S_{T i} \approx 0.10$) to the principal determinant of system behavior ($S_{T i}>0.60$). Thus, the Sobol' decomposition confirms that AI-labor coupling exhibits a stronger and more asymmetric dependence than AI-capital feedback, thereby substantiating the theoretical expectation of significant technological displacement and reinforcement within the expanded Lotka-Volterra framework. Fig. \ref{sobol_labor} shows the visualizations of  the Sobol indices for both $x^*$ and $y^*$ in the AI-labor subsystem. 

\newpage

\section{Conclusion and Policy Implications}
\label{s6}

This study has employed a coupled dynamical-systems framework to quantify the nonlinear interactions between artificial intelligence (AI) capital, physical capital, and labor in China. By treating AI as an independent factor of production \cite{lu2021} and modeling its co-evolution with traditional inputs via the Lotka–Volterra predator–prey structure, we have moved beyond static correlations to capture the feedback loops that characterize China's ongoing technological transition \cite{zhu2018}. The estimated parameters, equilibrium analysis, and global sensitivity results jointly yield a set of concrete, policy-relevant insights for managing AI-driven structural change.

Our core empirical finding is that AI capital in China acts as the ``prey" in its interactions with both physical capital and labor—stimulating the accumulation of the former and augmenting the wage bill of the latter \cite{he2019, ai1}, while itself being only weakly constrained by their feedback. This distinctive sign pattern, together with the stable nodal equilibria we identify, reflects the deliberate balancing act embedded in China's current AI governance. On one hand, industrial policies actively promote AI adoption to modernize traditional sectors \cite{huang2024}; on the other, regulatory safeguards (e.g., antitrust enforcement, employment-stability guidelines) prevent AI from wholly displacing existing capital stocks or labor demand \cite{jones}. The result is a policy-mediated dynamic that converges monotonically to a steady state, avoiding the volatile oscillations that could arise under less managed technological competition.

The sensitivity analysis further illuminates where policy leverage is greatest. For the AI–physical capital nexus, equilibrium outcomes depend meaningfully on both AI-side parameters and capital's own saturation dynamics, suggesting that industrial policies targeting either technological diffusion or traditional-sector capacity can effectively steer the long-run configuration. In contrast, the AI–labor subsystem is overwhelmingly driven by AI-related parameters, indicating that labor-market outcomes in an AI-intensive economy are primarily a function of how—and how fast—AI is deployed. Labor-side policies (e.g., wage floors, training programs) alone may have limited influence on the steady state unless they directly alter the AI-to-labor interaction strength.

Based on these findings, we propose three targeted implications for policymakers:

\begin{enumerate}
    \item \textbf{Calibrating AI Promotion Policies:} The weak negative feedback from physical capital to AI suggests that there remains policy space to accelerate AI investment without immediately triggering strong counter-pressures from traditional industries. However, the sensitivity results caution that once AI-labor coupling intensifies, labor-market outcomes become highly sensitive to AI-driven parameters, calling for forward-looking labor-market planning alongside AI stimulus.

    \item \textbf{Managing Structural Transition:} The stable-node property implies that the current policy mix succeeds in avoiding disruptive cycles, but it also points to structural rigidities that may slow optimal resource reallocation. Policymakers could consider gradually relaxing select constraints (e.g., sectoral entry barriers, labor-mobility restrictions) to allow for smoother adjustment while monitoring the system's stability margins.

    \item \textbf{Prioritizing Intervention Channels:} The asymmetric sensitivity across subsystems highlights that policy effectiveness depends on the target relationship. Industrial rebalancing between AI and physical capital requires a dual-track approach, whereas labor-market resilience in the face of AI hinges disproportionately on shaping AI's deployment patterns—for instance, through incentives for human-AI collaboration rather than full automation.
\end{enumerate}

In sum, this study provides a quantitative, system-aware foundation for designing AI policies that aim to harness growth potential while maintaining socioeconomic stability. The Lotka–Volterra framework, though parsimonious, captures essential feedback mechanisms often omitted in conventional policy models \cite{Lee2005, xia2022}. Future research should extend this approach to a three-factor model encompassing simultaneous AI–capital–labor interactions, incorporate spatial or sectoral heterogeneity \cite{huang2024}, and evaluate policy shocks through dynamical simulations. Such advancements would further enhance this tool's utility for evidence-based technological economic governance, helping policymakers navigate the complex trade-offs inherent in the age of artificial intelligence.





\end{document}